\documentclass[journal=jacsat,manuscript=article]{achemso}
\usepackage{graphicx}%
\usepackage{xcolor}
\usepackage{bm}
\usepackage{amsmath, amsthm, amssymb}
\usepackage{braket}
\renewcommand\Im{\operatorname{Im}}

\author{Julen Iba\~{n}ez-Azpiroz}
\email{j.azpiroz@fz-juelich.de}
\phone{+49 2461 61 8863}
\affiliation[FZJ]
{Peter Gr\"unberg Institute and Institute for Advanced Simulation, Forschungszentrum
J\"ulich \& JARA, D-52425 J\"ulich, Germany}
\author{Manuel dos Santos Dias}
\affiliation[FZJ]
{Peter Gr\"unberg Institute and Institute for Advanced Simulation, Forschungszentrum
J\"ulich \& JARA, D-52425 J\"ulich, Germany}
\author{Stefan Bl\"ugel}
\affiliation[FZJ]
{Peter Gr\"unberg Institute and Institute for Advanced Simulation, Forschungszentrum
J\"ulich \& JARA, D-52425 J\"ulich, Germany}
\author{Samir Lounis}
\affiliation[FZJ]
{Peter Gr\"unberg Institute and Institute for Advanced Simulation, Forschungszentrum
J\"ulich \& JARA, D-52425 J\"ulich, Germany}

\title[]
  {Zero-point spin-fluctuations of single adatoms}

\begin{document}




\begin{abstract}

Stabilizing the magnetic signal of single adatoms 
is a crucial step towards their successful usage in widespread 
technological applications such as high-density magnetic data  storage devices. 
The quantum mechanical nature of these tiny objects, however, 
introduces intrinsic zero-point spin-fluctuations 
that tend to destabilize the local magnetic moment of interest 
by dwindling the magnetic anisotropy potential barrier 
even at absolute zero temperature. 
Here, we elucidate the origins and quantify the effect of
the fundamental ingredients determining the magnitude of the fluctuations, 
namely the 
($i$) local magnetic moment, 
($ii$) spin-orbit coupling and ($iii$) electron-hole Stoner excitations.
Based on a systematic first-principles study of 3d and 4d adatoms, 
we demonstrate that the transverse contribution of the fluctuations 
is comparable in size to the magnetic moment itself, leading to a
remarkable  $\gtrsim$50$\%$ reduction of the magnetic anisotropy energy.
Our analysis gives rise to a comprehensible diagram relating the 
fluctuation magnitude to characteristic features of adatoms,
providing practical guidelines for designing 
magnetically stable nanomagnets with minimal quantum fluctuations.

\end{abstract}
\maketitle

Understanding electron spin dynamics and spin relaxation phenomena
of nanomagnets is of capital importance due to both, fundamental motivations 
and potential technological applications, for instance, in the context of 
magnetic data storage in the atomic limit.
If the hitherto ever-growing trend of magnetic storage density is to be mantained
in the future, the magnetic building blocks need to be shrinked to 
the size of just a handful of atoms.  
At this scale, however, quantum mechanics 
pose a serious threat to the stability of the local magnetic moment associated to the nanomagnets, 
which can be easily perturbed by 
interactions with their environment~\cite{hirjibehedin_large_2007,PhysRevLett.102.257203,prl-itinerant}.
In this regard, the stability of the magnetic signal
depends crucially on the so-called magnetic anisotropy
energy (MAE), an energy barrier generated by 
the spin-orbit coupling (SOC) that protects and stabilizes 
the direction of the local magnetic moment against possible fluctuations of the spin,
\textit{e.g.} of thermal origin.

Among nanomagnets, single magnetic adatoms represent the smallest 
possible magnetic unit, thus motivating an intense search for elements
that exhibit large and stable magnetic moments
when deposited on a substrate.
Early theoretical simulations 
based on density functional theory 
(DFT)~\cite{oswald_giant_1986,wildberger_magnetic_1995,lang_local_1994}  
boosted this search and created a huge enthusiasm in the field
by predicting gigantic local magnetic moments of 
diverse transition-metal (TM) adatoms, including 4d and 5d elements that
are nominally nonmagnetic in bulk.
The corresponding experimental scenario, in turn, 
is notably rich, complex and challenging. 
On one hand,
X-ray magnetic circular dichroism (XMCD) and inelastic scanning tunneling spectroscopy (ISTS)  
assert that several 3d TM adatoms can possess a substantial MAE of few meV
(see \textit{e.g.} refs~\cite{gambardella_giant_2003,rau_reaching_2014,PhysRevLett.108.256811,
heinrich_single-atom_2004,hirjibehedin_spin_2006,
heinrich_tuning_2015,PhysRevLett.111.157204,PhysRevLett.114.106807}).
On the other hand, and in remarkable contrast, the very same adatoms  
behave as paramagnetic entities 
when measured by means of spin-polarized scanning tunneling microscopy (SP-STM) 
(see \textit{e.g.} 
refs~\cite{meier_revealing_2008,khajetoorians_atom-by-atom_2012,zhou_strength_2010}),
implying the existence of a mechanism that destroys 
the magnetic stability locally.
Going one step further, the case of 4d and 5d adatoms is even more striking, 
given that they have so far exhibited 
no clear magnetic signal even when subjected to the static magnetic field of an XMCD experiment, 
in notorious disagreement with theoretical predictions~\cite{honolka_absence_2007}.

\begin{figure}[b]
\includegraphics[width=0.45\textwidth]{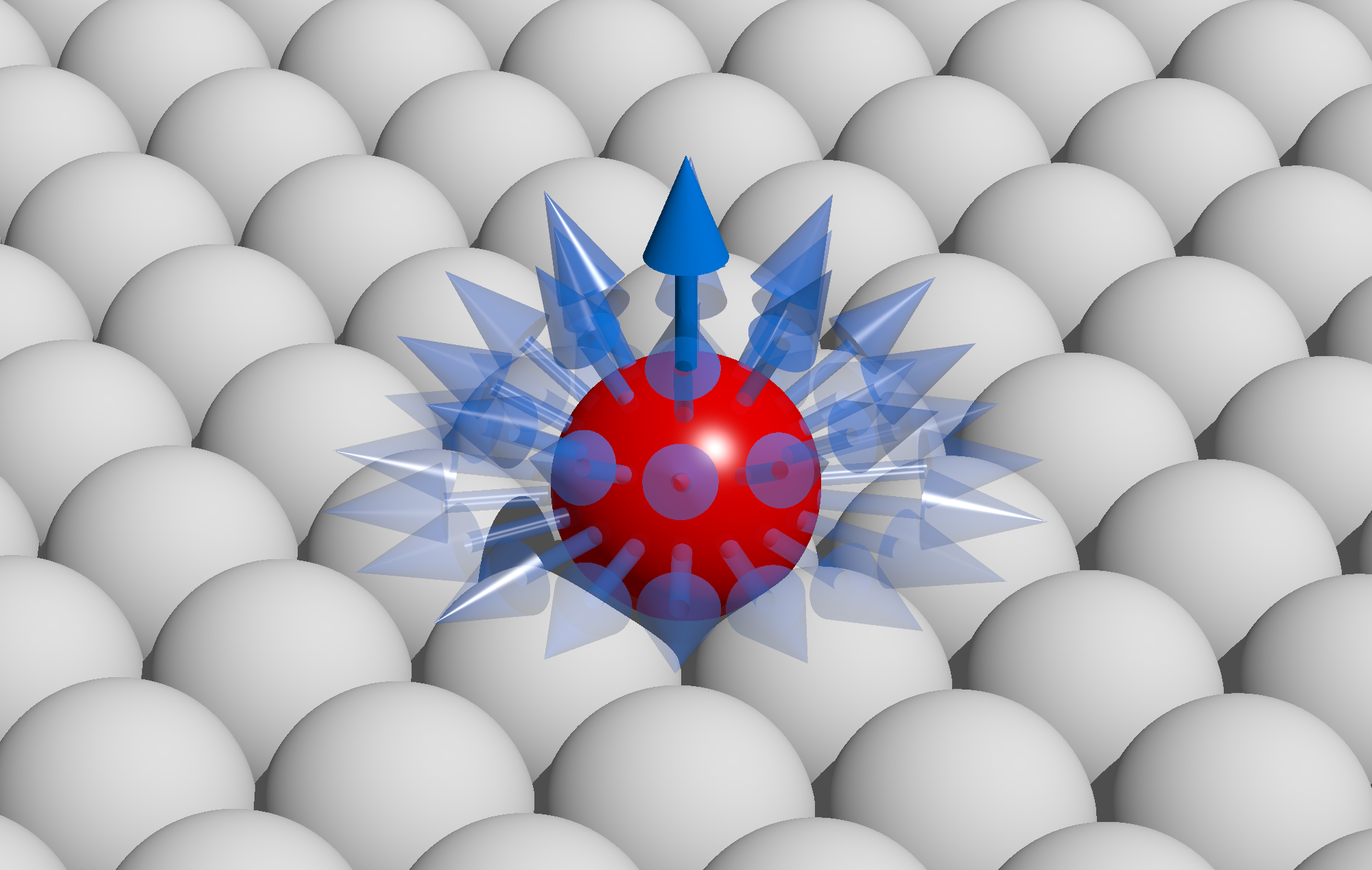}
\caption{Schematic illustration of the spin-fluctuations (blue arrows) of an adatom (red ball)
supported on a surface (grey balls).
}
\label{fig:fluc-image}
\end{figure}

In view of this scenario,
a central question arises: 
what is the mechanism leading to the apparent instability of the  
magnetic moment of an adatom?
In order to address this issue, 
here we investigate a key dynamical aspect of single adatoms 
that has not been hithertho considered, 
namely the contribution of zero-point spin-fluctuations (ZPSF)
(see Fig. \ref{fig:fluc-image} for a schematic illustration).
These are quantum fluctuations present even at absolute zero temperature 
that can crucially affect the magnetic properties 
of itinerant electron magnets~\cite{aguayo_why_2004,ortenzi_accounting_2012}. 
Here, we elucidate their origin and quantify their impact on the magnetic stability 
of the series of 3d and 4d TM adatoms deposited on metallic substrates.
Remarkably, our first principles investigation reveals that the transverse contribution to the ZPSF 
is  of the order of the local magnetic moment itself, an astonishingly large value
that has profound effects on the MAE, which can be reduced by more than
$50\%$ with respect to the static value calculated by standard DFT.   
In order to minimize their destabilizing effect,
we pinpoint the three fundamental ingredients that determine the magnitude of the ZPSF, 
namely the ($i$) local magnetic moment, 
($ii$) SOC and ($iii$) electron-hole Stoner excitations.
Based on our findings, 
we develop a simple diagram where the ZPSF of an arbitrary 
adatom are classified according to the aforementioned factors, 
offering practical guidelines for stabilizing 
robust magnetic properties against fluctuations.

The ZPSF are formally given by the fluctuation-dissipation theorem~\cite{PhysRev.83.34},
which relates the variance of the spin-fluctuations, $\xi^{2}$, 
to the imaginary part of the enhanced spin-susceptibility, 
$\Im \chi(\omega)$.
As it turns out, the ZPSF are predominantly determined by the 
transverse contribution (see Supporting Information), which is given by~\cite{PhysRev.83.34,book_vignale}
\begin{equation}
\label{eq:fd-theorem}
\xi_{\perp}^{2}= -\dfrac{1}{\pi} \int_{0}^{\infty} d\omega \Im \chi_{\perp}(\omega).
\end{equation}
Interestingly, Eq. \ref{eq:fd-theorem} establishes a deep and insightful connection between the
internal fluctuations and the response properties of the system, but its
calculation is hard in practice,  and no systematic first-principles 
study of the above quantity has been performed so far.

\begin{figure}[t]
\includegraphics[width=0.75\textwidth]{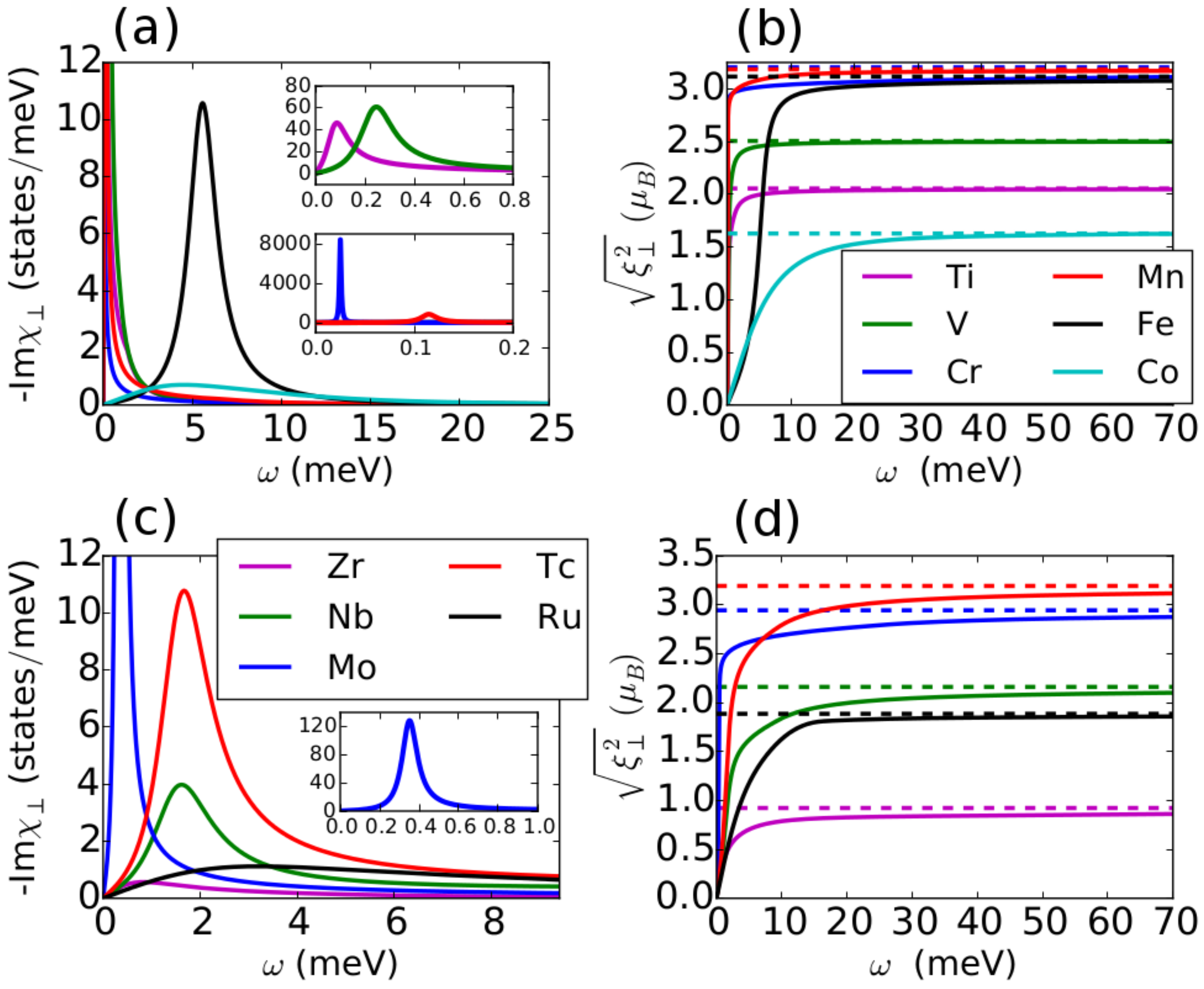}
\caption{(a) and (c) Density of transverse spin-excitations as given by
$\Im \chi_{\perp}(\omega)$ for selected 3d and 4d adatoms on Ag(100), respectively. 
Insets adress different frequency regions where various resonance frequencies are located.
(b) and (d) Calculated magnitude of the 
mean value of transverse ZPSF for the adatoms
considered in (a) and (c), respectively.
Solid lines depict the evolution of the value
as a function of the upper boundary of the integral of Eq. \ref{eq:fd-theorem}, while the
horizontal dashed lines represent the converged value.
}
\label{fig:imchi-xi}
\end{figure}

Let us begin our analysis by discussing Fig. \ref{fig:imchi-xi}, where 
the calculated spin-excitation and spin-fluctuation spectra
associated to TM adatoms deposited on Ag(100) are displayed
(see Supporting Information regarding technical 
details on the formalism and calculations). 
First, we  focus on Figs.  \ref{fig:imchi-xi}(a,c), which 
illustrate the calculated $\Im\chi_{\perp}(\omega)$ for 3d and 4d adatoms, respectively. 
These figures reveal
the existence of a large peak in the meV range for all adatoms, 
corresponding to a spin-excitation.
Noteworthy, the location of the resonance frequency 
is proportional to the MAE, which in turn is determined by the SOC~\cite{dias_relativistic_2015}.
Fe shows the largest resonance frequency with $\omega_{0}\sim6$ meV, 
while for the rest of adatoms we find $\omega_{0}\lesssim 4$ meV; in the extreme cases of
Ti, V, Mn, Cr and Mo, the spin-excitation is found at very small frequencies, $\omega_{0}\lesssim 0.5$ meV,
impliying that the net effect of SOC is extremely weak in these adatoms. 
A further feature revealed by Figs. \ref{fig:imchi-xi}(a,c)
is the width of the spin-excitation peak, 
which is linked to the amount of electron-hole
Stoner excitations near the Fermi level~\cite{PhysRevB.91.104420}.
Our calculations show that Ru and, to some extent also Nb and Tc, possess large widths as compared to the rest of adatoms, specially
those that peak below 1 meV.

Next, we analyze the magnitude of the transverse ZPSF (Eq. \ref{eq:fd-theorem}),
as illustrated in Figs. \ref{fig:imchi-xi}(b,d)  
for 3d and 4d adatoms on Ag(100), respectively; 
the solid lines depict the evolution 
of the mean value, $\sqrt{\xi_{\perp}^{2}}$, as a function of the upper 
boundary of the frequency integral,
while the converged value
is denoted by the horizontal dashed lines (see Supporting Information). 
The most important message of  these figures is that  
$\sqrt{\xi_{\perp}^{2}}$ is of the order of the Bohr magneton;
the elements with largest values are Cr, Mn, Fe and Tc, 
which have $\sqrt{\xi_{\perp}^{2}}\sim3$ $\mu_{B}$,
while in the case of Zr and Co this value is drastically reduced by more than 50 $\%$.
Our calculations further demonstrate that the main contribution to the integral of 
Eq. \ref{eq:fd-theorem} comes from the spin-excitation peak in the meV region,
which represents between 70 $\%$ and $\sim$100 $\%$ of the total depending on the adatom.

\begin{figure}[t]
\includegraphics[width=0.75\textwidth]{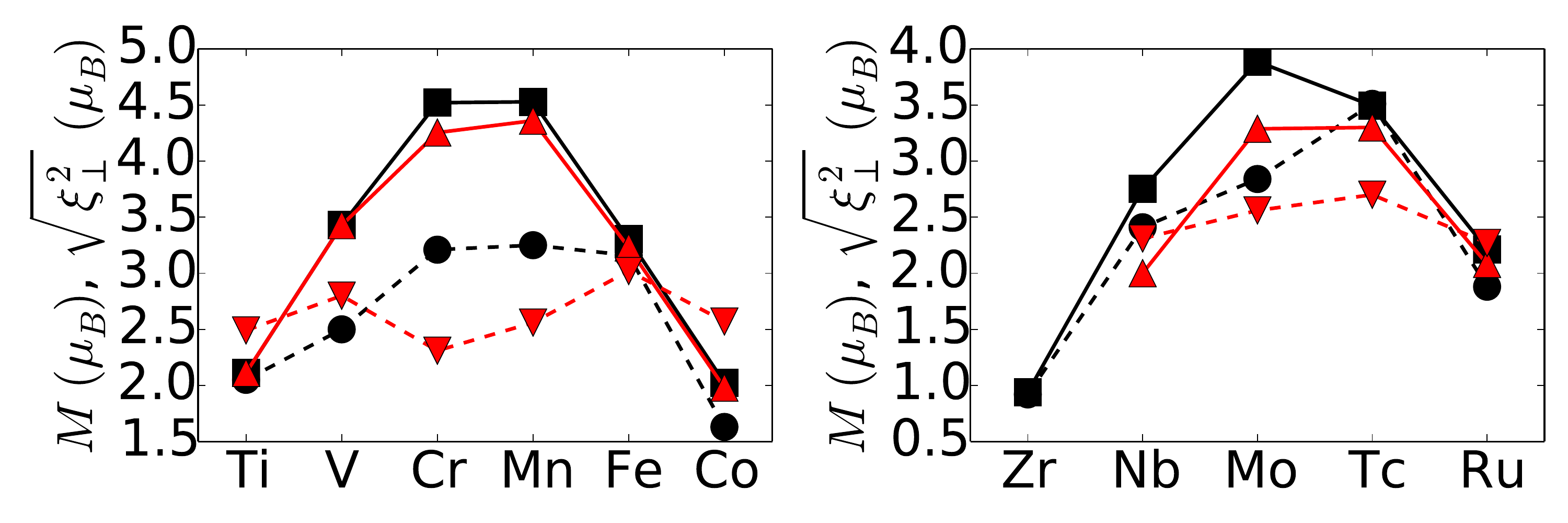}
\caption{Local magnetic moment (solid lines) and mean value of the transverse 
ZPSF (dashed lines) for 3d (left pannel) and 4d (right pannel) adatoms. $M$ ($\sqrt{\xi^{2}_{\perp}}$)
is denoted by black squares (black circles) and red upward triangles (red downward triangles)
for adatoms deposited on Ag(100) and Cu(111), respectively. 
}
\label{fig:M0-xi}
\end{figure}

It is instructive to compare the  magnitude of the transverse ZPSF with the
local magnetic moments of the adatoms, which we denote as $M$. 
This is done in Fig. \ref{fig:M0-xi},
where it is demonstrated that $\sqrt{\xi^{2}_{\perp}}$ (black circles) 
represents always an appreciable fraction of $M$ (black squares); 
we find $\sqrt{\xi^{2}_{\perp}}\sim M/2$ for Mn and Cr, while  
$\sqrt{\xi^{2}_{\perp}}\sim M$ for Ti, Fe, Zr, Nb and Ru,  
implying that the latter suffer from very strong deviations
of the direction of the local magnetic moment.
It is also noteworthy that both $M$ and $\sqrt{\xi^{2}_{\perp}}$ 
follow the evolution dictated by Hund's rules, whereby the
adatoms with nearly half-filled d-shells have largest values: case of Cr and Mn among 3d, 
and Mo and Tc among 4d.
This trend is clearly fulfilled in the case of the magnetic moment, 
as reported in previous works~\cite{wildberger_magnetic_1995,oswald_giant_1986},
whereas the evolution of the fluctuation magnitude presents some exceptions, such as 
the case of Fe and Tc.
A further feature revealed by Fig. \ref{fig:M0-xi} is that the 
spin-fluctuation-to-magnetization ratio (SFMR) is overall larger
in 4d adatoms than in 3d.
As a final remark, we note that Mo has by far the lowest 
SFMR among 4d elements: interestingly,   
it is the only 4d adatom deposited on Ag(100) 
that exhibits an experimentally detectable magnetic signal~\cite{mo-ag100}.

We have extended the above analysis to the same set of adatoms deposited on Cu(111),
which exhibit essentially the same features 
as on Ag(100) (see red triangles in Fig. \ref{fig:M0-xi}).
The only mentionable difference is that the SFMR of Cr and Mn 
is somewhat lower than what is expected from the trend of Fig. \ref{fig:M0-xi};
the origin of this feature will be discussed in the next subsection.
Our \textit{ab initio} investigation  has therefore exposed an important general property 
of adatoms, namely that the 
magnitude of their transverse
ZPSF is of the order of their local magnetic moment.

\begin{figure}[t]
\includegraphics[width=0.7\textwidth]{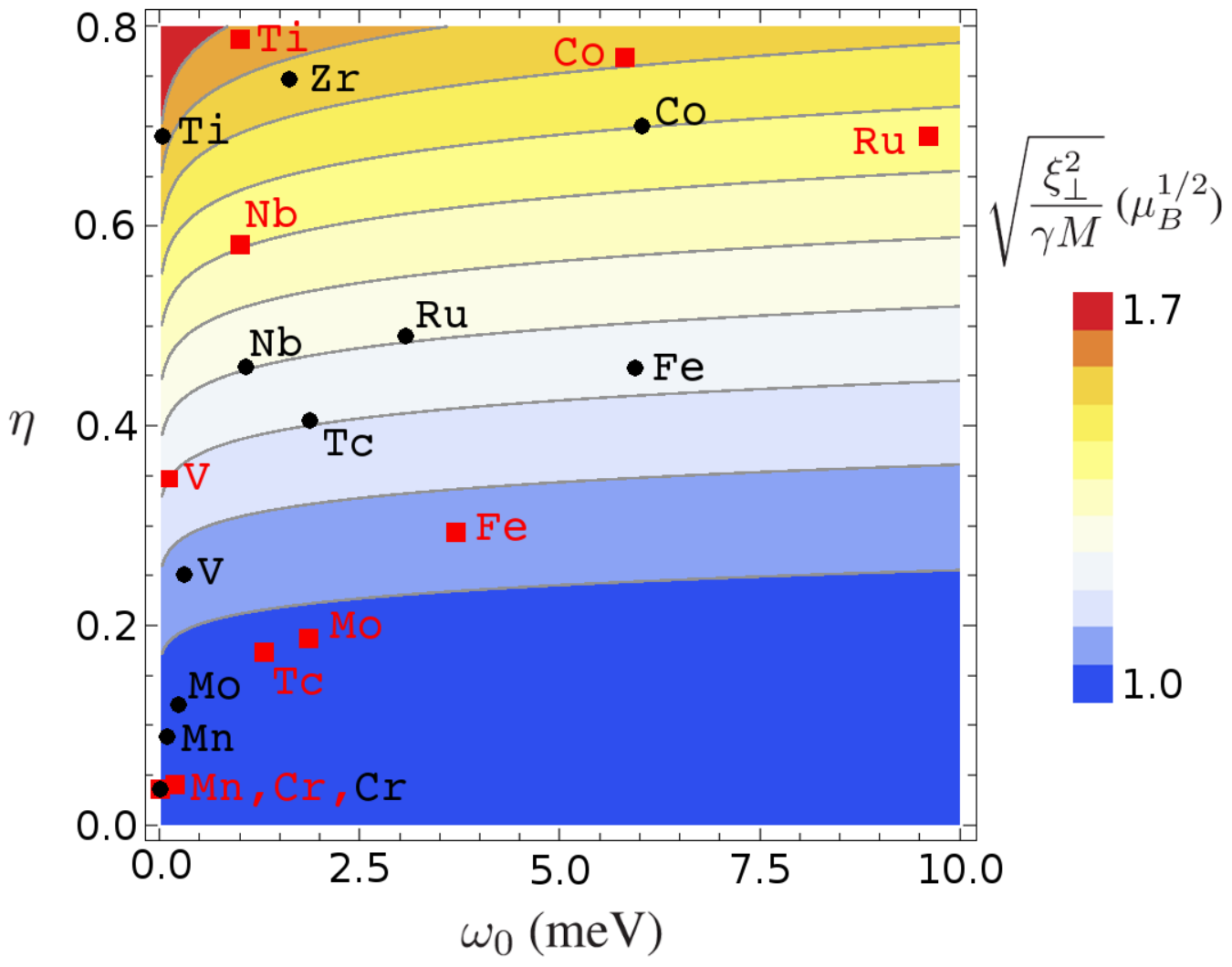}
\caption{The background shows a 2D plot of the 
SFMR as a function of the damping
and the resonance frequency in the LLG model (Eq. \ref{eq:LLG-xi}). 
Circles (black) and squares (red) denote various adatoms 
deposited on Ag(100) and Cu(111),
respectively, where the parameters $\eta$ and $\omega_{0}$ have been extracted from a fit to
the \textit{ab initio} spin-susceptibility.}
\label{fig:LLG-xi-eta-wres}
\end{figure}

We proceed now to identify the fundamental factors that determine the 
magnitude of the transverse ZPSF.
For such purpose, we consider  the 
Landau-Lifshitz-Gilbert  (LLG)  equation~\cite{gilbert_phenomenological_2004}.
This is widely employed for characterizing the spin-dynamics of macroscopic
magnetic systems, and its use has been recently legitimated also for
microscopic systems~\cite{PhysRevB.91.104420,dias_relativistic_2015},
as it allows to accurately reproduce the \textit{ab initio} calculations 
by extracting the relevant parameters. 
In the LLG model, the imaginary part of the transverse spin-susceptibility takes the form
of an skewed Lorentzian, \textit{i.e.} $\text{Im}\chi_{\pm}^{\text{LLG}}(\omega)=C
\cdot\eta\omega/\big((\omega-\omega_{0})^{2}+(\eta\omega_{0})^{2}\big)$,
with $C=M\gamma/2(1+\eta^{2})$. 
The parameters entering this model are
the Gilbert damping, $\eta$, which is proportional to the width of the spin-excitation peak and is
therefore dominated by Stoner excitations, and the resonance frequency,  
$\omega_{\text{res}}=\omega_{0}\sqrt{1+\eta^{2}}=\gamma B_{\text{eff}}/\sqrt{1+\eta^{2}}$,
where $\gamma$ is the gyromagnetic ratio
and $B_{\text{eff}}$ an effective magnetic field whose magnitude is determined by 
the strength of SOC; 
a detailed discussion can be found in, \textit{e.g.}, Ref.~\cite{dias_relativistic_2015}.

Within the LLG model, the integral of Eq. \ref{eq:fd-theorem} can be calculated analytically,
yielding the following expression for the transverse ZPSF:
\begin{equation}
\label{eq:LLG-xi}
\xi^{2}_{\text{LLG}}=
\dfrac{M\gamma}{\pi(1+\eta^{2})}
\Big(
\dfrac{\eta}{2} \log\dfrac{(x^{2}+\eta^{-2}+1)^{2}-(2x\eta^{-1})^{2}}{(\eta^{-2}+1)^{2}} + 
\vartheta(x,\eta)
\Big).
\end{equation}
Above, $\vartheta(x,\eta)=\arctan (x-\eta^{-1}) - 
\arctan (x+\eta^{-1}) + 2\arctan \eta^{-1}$
and $x=\omega_{c}/\eta\omega_{0}$,
with $\omega_{c}$ a 
cutoff frequency to be converged (see Supporting Information).

Eq. \ref{eq:LLG-xi} is very useful as it provides an interpretation
for the magnitude of the fluctuations in terms of the physical parameters of the LLG model, which
are in turn related to the electronic structure of the adatoms.
As anticipated in the introduction, three major ingredients come into play:
$M$, $\omega_{0}$ and $\eta$.
Eq. \ref{eq:LLG-xi} reveals that $\sqrt{\xi^{2}_{\text{LLG}}}$ is directly proportional to $\sqrt{M}$, 
which, apart from accounting for 
the rough proportionality displayed by our calculations (see Fig. \ref{fig:M0-xi}),
it indicates that fluctuations are 
relatively weaker for adatoms with large magnetic moments, \textit{i.e.}
the nearly half-filled d-shell elements.  
Interestingly, the simple relationship of Eq. \ref{eq:LLG-xi}  allows to analyze the SFMR,  
$\sqrt{\xi^{2}_{\text{LLG}}/\gamma M}$, 
as a function of $\eta$ and $\omega_{0}$ by setting the standard value 
for the gyromagnetic ratio $\gamma=2$; the resulting map is displayed in the background of 
Fig. \ref{fig:LLG-xi-eta-wres}.
This figure evidences that the SFMR
is critically enhanced by the damping, as $\sqrt{\xi^{2}_{\text{LLG}}/\gamma M}$ varies by almost  
70 $\%$ in the range of values considered for $\eta$.
This feature comes from the fact that
enhanced Stoner excitations give rise to a large dissipation of energy,
implying large fluctuations via the connection established by the fluctuation-dissipation
theorem. 
It is also noteworthy that 
when $\eta$ tends to zero,
$\sqrt{\xi^{2}_{\text{LLG}}/\gamma M}=1$ $\mu_{B}^{1/2}$, \textit{i.e.} 
Eq. \ref{eq:LLG-xi} reveals an
intrinsic contribution to the
transverse ZPSF present even in the limiting case of a spin-excitation with vanishing width. 
Finally, Fig. \ref{fig:LLG-xi-eta-wres} demonstrates that the transverse ZPSF are reduced for large
resonance frequencies, but, quite unexpectedly, one finds that 
the induced variation 
is much less important than with the damping.

We are now in position to perform a quantitative analysis of the \textit{ab initio}
transverse ZPSF in terms of the LLG model. For such purpose,
we have systematically extracted the parameters $\eta$ and  $\omega_{0}$ by fitting  
$\text{Im}\chi_{\pm}^{\text{LLG}}(\omega)$ to the
calculated spin-susceptibility (\textit{e.g.} Figs. \ref{fig:imchi-xi}(a,c))
for all adatoms.
This allows us to locate the position of each adatom on the map
of Fig. \ref{fig:LLG-xi-eta-wres}, 
as depicted by the circles and squares for the case of the Cu(111) and Ag(100) substrates, respectively.
The resulting distribution makes it clear that the origin of the 
large SFMR found in the case of 
Ti, Co, Ru and Nb on both substrates, as well as Fe and Zr on Ag(100),
is mainly due to the large damping factors of these adatoms, $\eta\gtrsim0.5$, while
this tendency is only slightly modified by the position of the resonance frequency.
On the opposite side, despite the small resonance frequency of 
Mn, Cr and Mo adatoms on both substrates, their extremely small damping, $\eta\lesssim 0.05$, 
makes $\sqrt{\xi^{2}_{\text{LLG}}/\gamma M}$ approach 
the intrinsic minimum value.

We note that the distribution of elements shown in Fig. \ref{fig:LLG-xi-eta-wres}
is a guideline to understand the behavior of the SFMR
in the context of the LLG model,
not an exhaustive representation of the \textit{ab initio} data
summarized in Fig. \ref{fig:M0-xi}.
For this reason, some details,  such as the larger SFMR of Cr and Mn on Cu(111) than on Ag(100)
(see Fig. \ref{fig:M0-xi}) are not accounted for by the distribution.
This particular feature can be attributed to
deviations of the gyromagnetic ratio 
from the standard value $\gamma=2$, which are larger 
in the case of Cr and Mn on Cu(111) than on Ag(100)~\cite{dias_relativistic_2015} (see Eq. \ref{eq:LLG-xi}).

Having provided an interpretation for the calculated magnitude of the transverse ZPSF 
in terms of the underlying physical parameters, we come now to
analyze its effects on the magnetic stability of the adatoms.
For such purpose, we estimate how the fluctuations affect the MAE,
as this quantity defines the strength of the magnetic easy axis. 
We note that the energy scale of the 
MAE for adatoms is meV~\cite{gambardella_giant_2003,dias_relativistic_2015},
which coincides with the energy of spin-excitations that give rise to 
the primary contribution to the transverse ZPSF (see Fig. \ref{fig:imchi-xi}).

\begin{figure}[t]
\includegraphics[width=0.75\textwidth]{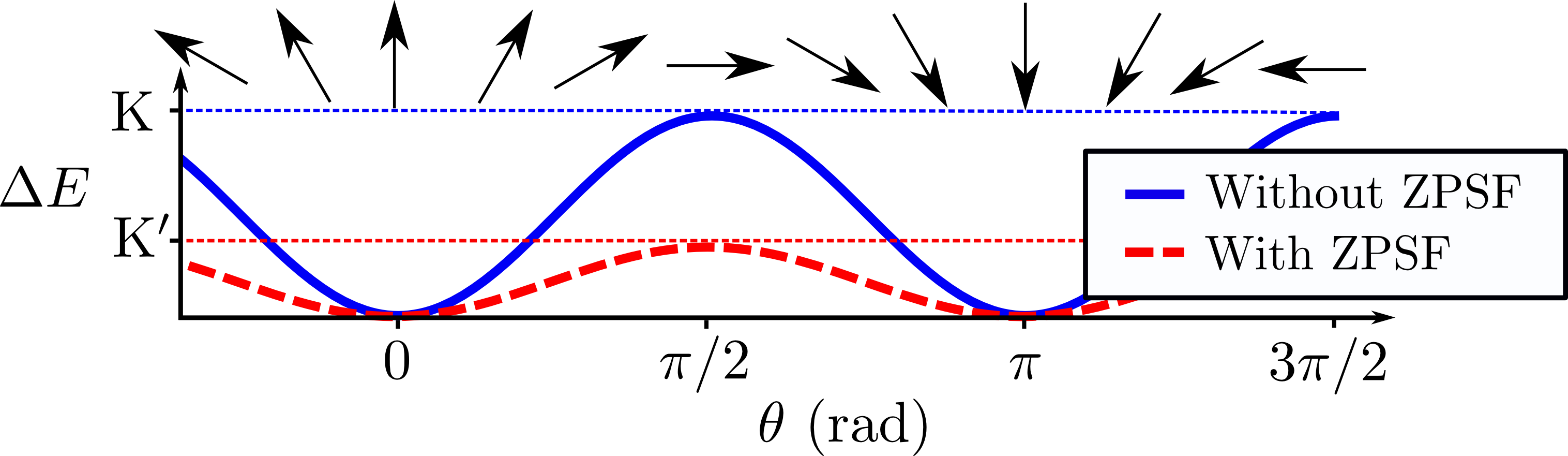}
\caption{Schematic illustration of the renormalization of the MAE induced by the
transverse ZPSF. 
$K$ and $K'$ respectively represent the static and renormalized anisotropy constant
(see Eq. \ref{eq:mae-xi}).
}
\label{fig:MAE}
\end{figure}
\begin{table}[t]
 \begin{tabular}{ c | c | c | c | c || c | c | c | c }
  \hline                       
                         & Cr & Mn & Fe & Co & Nb & Mo & Tc & Ru  \\ 
  \hline  
           \multicolumn{9}{c}{\hspace{1cm} On Ag(100)} \\
         \hline
  K (meV)                & 0.22 & $-$0.02 & $-$3.66 & $-$0.46 & 1.78 & 1.55 & $-$1.13 & $-$8.32 \\
K$'$ (meV)               & 0.05 & $-$0.01 & $-$0.12 & $-$0.08 & 0.16 & 0.35 &  0.01 & $-$0.99 \\
\hline  
           \multicolumn{9}{c}{\hspace{1cm}  On Cu(111)} \\
           \hline
  K (meV)                & 0.21 & 0.36 & $-$4.73 & $-$3.7 & 0.55 & 2.60 & 3.13 & $-$20.01 \\     
  K$'$ (meV)             & 0.07 & 0.11 & $-$0.32 & 1.12 & $-$0.05 & 0.46 & 0.44 & 2.66 \\    
         \hline  
    \end{tabular}
     \caption{Calculated anisotropy constants K and K$'$ for selected 3d and 4d 
    adatoms on Ag(100) and Cu(111).\label{table:MAE}}
        \end{table}

Let us consider the expression for the MAE for uniaxial systems, 
$E_{a}(\theta)=\text{K}(\textbf{M}\cdot\hat{\textbf{e}}_{z})^{2}/\textbf{M}^{2}=
\text{K}\cos^{2}\theta$,
where K is the so-called anisotropy constant.
Consequently, the energy barrier between the magnetic moment pointing 
along the $z$ axis ($\theta=0$) and a perpendicular axis ($\theta=\pi/2$)
is simply given by
$\Delta E = E_{a}(\theta=0)-E_{a}(\theta=\pi/2)=\text{K}$,
as schematically illustrated in Fig. \ref{fig:MAE}.
In the spirit of the spin-fluctuation theory of Moriya~\cite{moriya-book}, 
we now let magentic moment to fluctuate around its equilibrium value,
\textit{i.e.} $\textbf{M}^{2}\rightarrow 
 \left( 
M\hat{\textbf{e}}_{z}+\sum_{\perp}\boldsymbol{\xi}_{\perp}
\right)^{2} $.
Introducing this term into the definition of $E_{a}(\theta)$,
we obtain a renormalized expression for the MAE, \textit{i.e.}
$E_{a}(\theta,\xi^{2}_{\perp})=\text{K}(M^{2}\cos^{2}\theta+\xi^{2}_{\perp}\sin^{2}\theta)/(M^{2}+2\xi^{2}_{\perp})$.
Noteworthy, this implies that the energy barrier gets effectively reduced by the transverse ZPSF,
\begin{equation}
\label{eq:mae-xi}
\Delta E(\xi^{2}_{\perp})= 
\text{K}\left(
1-\dfrac{3\xi^{2}_{\perp}}{M^{2}+2\xi^{2}_{\perp}}\right)\equiv \text{K}',
\end{equation}
which is characterized by a modified anisotropy 
constant, K$'$, as schematically illustrated in Fig. \ref{fig:MAE}.

In Table \ref{table:MAE} we have listed the calculated values for both, 
the anisotropy constant K, which we have evaluated by band energy differences following the magnetic 
force theorem~\cite{PhysRevB.32.2115}, and the reduced constant K$'$, calculated using the
values of the magnetic moments and transverse ZPSF of the adatoms.
Our calculations reveal a strong and generalized 
reduction of the MAE, which goes
much beyond previous estimates 
that did not take  into account fluctuation effects~\cite{dias_relativistic_2015}. 
For the extreme cases where the ZPSF are larger than $M$ (mostly 4d adatoms),
the direction of the local magnetic moment gets completely destabilized, 
as reflected by the sign change in the renormalized MAE. 
It is particularly noteworthy that in the well-studied  Fe on Cu(111) system, the
ZPSF considerably fix the disagreement between the 
experimental MAE ($\sim-1$ meV~\cite{prl-itinerant,PhysRevB.91.235426})
and the theoretical one, which is reduced from K$=-4.73$ meV to K$'=-0.32$ meV.
Aside from the numerics, this feature demonstrates that the renormalization of the MAE
due to the ZPSF has the correct order of magnitude.
Finally, for the 4d TM elements Ru and Nb on Ag(100), 
where experiments remarkably fail to measure 
sizable magnetic moments~\cite{honolka_absence_2007,schafer-nb},
the reduction of the MAE predicted by our calculations 
is notably large ($\gtrsim90\%$), 
indicating that the ZPSF can also play a major role
in the absence of stable magnetism observed experimentally.

We have performed a systematic investigation of the impact of transverse ZPSF on the
magnetic properties of TM adatoms deposited on metallic substrates.
The magnitude of the fluctuations has been accessed 
via the fluctuation-dissipation theorem, employing 
the spin-susceptibility of the adatoms 
calculated \textit{ab initio} within TDDFT.
Our analysis has revealed that the transverse ZPSF represent 
an appreciable fraction of the local magnetic moment
of the adatoms, therefore strongly affecting their spin-dynamics. 
We have identified the nature of the three main ingredients that determine the magnitude
of the fluctuations, namely the $(i)$ SOC, $(ii)$ Stoner excitations and  $(iii)$ 
local magnetic moment itself, providing guidance for 
future search for systems with high stability of the magnetic signal.
Additionally, we have shown that spin-fluctuations can strongly reduce 
the MAE of the adatoms, therefore affecting their magnetic stability and
sheding light on the controversies between 
previous theoretical calculations and experimental observations.

Based on our investigation, we envision that 
the ideal adatom for technological applications, \textit{i.e.} the 
magnetically most stable one, 
should not only posses a large magnetic moment of possibly $4-5$ $\mu_{B}$, but 
also a large MAE and, above all, 
a very small damping 
(\textit{i.e.} virtually no Stoner excitations near the Fermi level) 
in order to drive
the fluctuations to their intrinsic quantum minimum. 
Within this scenario, the 
nearly half-filled elements Cr, Mn and Mo emerge as 
the most promising candidates given that 
they strongly fulfill the first and third conditions; however,
they exhibit a way too small MAE for real applications
when deposited on the Ag(100) and Cu(111) substrates here investigated.
The challenge resides therefore in tuning the anisotropy to its maximum value
by modifying either the nature of the substrate or the environment of the adatom
(\textit{e.g.} nanoislands, surface states, etc.),
a subject of intense ongoing investigation 
(see \textit{e.g.} refs~\cite{rau_reaching_2014,heinrich_tuning_2015,pbge111-spin-flip,prl-mae}).

The conclusions drawn by the 
single-adatom case studied in this work
settle as a solid background for tackling 
more complex nanomagnets constituted by several 
exchange-coupled magnetic adatoms.
This playground is particularly attractive given that, 
contrary to the single-adatom case, 
several SP-STM experiments have reported 
an unambiguous observation of 
stable magnetic moments
in this type of composite system
(see \textit{e.g.} refs~\cite{loth_bistability_2012,khajetoorians_current-driven_2013}),
strongly suggesting that local fluctuation effects decrease with increasing cluster size.
Interestingly, this feature is in qualitative 
accordance with one of our main findings in this work, 
namely that the SFMR associated to the transverse ZPSF 
becomes smaller the larger the magnetic moment gets.
This general trend, however, is likely to be modified by   
characteristic features associated to clusters,
such as the appearance of optical modes in the spin-excitation spectrum 
(\textit{i.e.} additional peaks in $\Im \chi_{\perp}(\omega)$, 
see \textit{e.g.} refs~\cite{lounis_dynamical_2010,lounis_theory_2011}) 
that would contribute to the ZPSF as
seen from Eq. \ref{eq:fd-theorem}.
On top of this, 
the impact of the ZPSF on the magnetic interactions among different adatoms
remains to be fully explored and clarified.
We believe that these future steps, alongside with the ones already taken in this work, 
can crucially contribute to current 
research efforts for understanding the spin dynamic phenomena of nanomagnets
and their stability.

\section*{Acknowledgments}
We thank M.C.T.D. M\"uller, E.  \c{S}a\c{s}io\u{g}lu,  
B. Schweflinghaus, F. Guimar\~{a}es, A. Eiguren and B. Baxevanis for 
many useful comments and discussions.
This work has been funded by the Helmholtz Gemeinschaft Deutscher-Young Investigators Group Program No. VH-NG-717 
(Functional Nanoscale Structure and Probe Simulation Laboratory),
the Impuls und Vernetzungsfonds der Helmholtz-Gemeinschaft Postdoc Programme
and  the European Research Council (ERC) under the European Union's Horizon 2020 research and innovation programme (ERC-consolidator grant 681405 — DYNASORE)


\bibliography{biblio}


\end{document}